\def\zqq{Z^0\rightarrow q\:{\overline q}}

\documentstyle[12pt,epsf]{article}

\begin{document}


\pagestyle{empty}

\renewcommand{\thefootnote}{\fnsymbol{footnote}}
                                                  

\begin{flushright}
{\small
SLAC--PUB--8206\\
July 1999\\}
\end{flushright}
                
\begin{center}
{\bf\Large
 Measurement of the $B^+$ and $B^0$ Lifetimes using Topological 
 Vertexing at SLD \footnote{Work supported in part by the
Department of Energy contract  DE--AC03--76SF00515.}}

\bigskip

The SLD Collaboration$^{**}$
\smallskip

Stanford Linear Accelerator Center, \\
Stanford University, Stanford, CA 94309\\
\medskip

\vspace{2.5cm}

{\bf\large
Abstract }
\end{center}
\noindent
   The lifetimes of $B^+$ and $B^0$ mesons have been measured using
   the entire sample of 550,000 hadronic $Z^0$ decays collected by the
   SLD experiment at the SLC between 1993 and 1998.
   In this paper, we describe the inclusive analysis of the 350,000 
   hadronic $Z^0$ decays collected in 1997-98 with the upgraded SLD
   vertex detector. In this data period,  
   a high statistics sample of 30903 (20731) charged 
   (neutral) vertices with good charge purity is obtained.
   The charge purity is enhanced by using the vertex mass, the SLC
   electron beam polarization (73\% for 1997-8)
   and an opposite hemisphere jet charge technique. Combining the
   results of this data sample with the results from the earlier
   data yield the following preliminary values:
   $\tau_{B^+}=1.623\pm0.020($stat$)\pm0.034$(syst) ps, 
   $\tau_{B^0}=1.589\pm0.021($stat$)\pm0.043$(syst) ps, 
   $\tau_{B^+}/\tau_{B^0} = 1.037\pm^{0.025}_{0.024}($stat$)
   \pm0.024$(syst). 

\vspace{1.5cm}

\begin{center}

{\sl Paper Contributed to the International Europhysics 
     Conference on High Energy Physics, July 15-21, 1999, Tampere,
     Finland, Ref. 5-477, and to the XIXth International Symposium on
     Lepton and Photon Interactions, August 9-14, 1999, Stanford,
     California, USA.}

\end{center}
\vfill

\normalsize

\pagebreak
\pagestyle{plain}

\pagebreak

 The spectator model predicts that the lifetime of a 
 heavy hadron depends
 upon the properties of the constituent weakly decaying heavy quark $Q$ and 
 is independent of the remaining, or spectator, quarks in the hadron.
 This model fails for the charm hadron system for which the
 lifetime hierarchy 
 $\tau_{D^+} \sim 2.3\tau_{D_s^+} \sim 2.5\tau_{D^0} 
 \sim 5\tau_{\Lambda_c^+}$ is observed.
 Since corrections to the spectator model are predicted to scale with
  $1/m_Q^2$ the $B$ meson lifetimes are expected to
  differ by less than 10\% \cite{bigi}. 
 Hence a measurement of the $B^+$ and $B^0$ lifetimes 
 provides a test of this prediction.
In addition, specific $B$ meson lifetimes are needed for many
 important measurements, e.g. to determine the  
 element $V_{cb}$ of the CKM matrix. 

 An analysis has been reported using the 
 1993-5 data sample of 150,000 hadronic 
 $Z^0$ decays collected with the original CCD vertex detector 
 (VXD2) as well as with the first 50,000 hadronic
 $Z^0$ decays collected in 1996 using the upgraded vertex detector (VXD3)
 \cite{vxd3}. In this paper we describe the
 analysis of the additional 350,000 hadronic $Z^0$ decays collected in
 the 1997-1998  run by the SLD detector at the SLC, and combine these
 results with those from the previous work\cite{9396results}.
 The excellent 3-D vertexing capabilities of SLD are exploited 
 with an inclusive topological vertexing technique \cite{zvnim} to
 identify $B$ hadron vertices
 produced in hadronic $Z^0$ decays with high efficiency
 (This inclusive technique has the advantage of very efficient
  $B$ vertex reconstruction since most $B$ decays are used).
 The decay length is measured using the
reconstructed vertex location while the $B$ hadron charge is determined from
the total charge of the tracks associated with the vertex. Knowledge
 of the average energy of $B$-hadrons produced in $Z^0$ decays, in
 conjuction with the decay length, lets one infer the $B$ lifetime.

The components of the SLD 
 utilized by this analysis are the Central
Drift Chamber (CDC)\cite{rbrb}
for charged track reconstruction and momentum measurement and the CCD pixel
Vertex Detector (VXD)\cite{vxd3,rbrb}
 for precise position measurements near the interaction
point. These systems are immersed in the 0.6 T field of the SLD solenoid.
Charged tracks reconstructed in the CDC are linked with pixel clusters in the
VXD by extrapolating each track and selecting the best set of associated
clusters\cite{rbrb}.
For VXD3 the track impact parameter resolutions at high momentum
are 9~$\mu$m and 11~$\mu$m in the $r\phi$ and $rz$ projections
respectively ($z$ points along the beam direction),
while multiple scattering contributions are
$33 \,\mu$m~$/(p\,{\rm sin}^{3/2}\theta)$ in both projections (where the
momentum $p$ is expressed in GeV/c).

The centroid of the micron-sized SLC Interaction Point (IP) in the $r\phi$
plane is reconstructed with a
measured precision of $\sigma_{IP} = (4\pm2)\, \mu$m using tracks in sets of
$\sim30$ sequential hadronic $Z^0$ decays. The median $z$ position of tracks
at their point of closest approach to the IP in the $r\phi$ plane is used to
determine the $z$ position of the $Z^0$ primary vertex on an event-by-event
basis.  A precision of $\sim30\,\mu$m on this
quantity is estimated using $Z^0 \rightarrow b\overline{b}$
Monte Carlo simulation for VXD3.

 The simulated $\zqq$ events are generated using JETSET 7.4 \cite{jetset}.
 The $B$ meson decays are simulated using the CLEO $B$ decay model
 \cite{CLEO-QQ} tuned to reproduce the spectra and
 multiplicities
 of charmed hadrons, pions, kaons, protons and leptons as measured at the
 $\Upsilon$(4S) by ARGUS and CLEO \cite{argcl}.
 The branching fractions
 of the charm hadrons are tuned to the existing measurements
 \cite{pdg98}. The $B$ mesons and baryons are generated
 with lifetimes of $\tau_{B^+}=1.64$ ps, $\tau_{B^0}=1.55$ ps,
 $\tau_{B^0_s}=1.57$ ps, and $\tau_{\Lambda_b}=1.22$ ps.
 The $b$-quark fragmentation follows the Peterson {\em et al.} parameterization
 \cite{Peterson}; the mean value of the fragmentation function in the 
 MC generation was $0.698$.
 The SLD detector is simulated using GEANT 3.21 \cite{geant}.

  Hadronic $Z^0$ event selection requires at least 7
 CDC tracks  which pass within 5~cm of the IP in $z$ at the point
 of closest approach to the beam and which have
 momentum transverse to the beam direction $p_T>$200~MeV/$c$.
 The sum of the energy of the charged tracks passing these cuts
 must be greater than 18~GeV.
 These requirements remove background from $Z^0 \to
 l^+ l^-$ events and two-photon interactions. In addition, the
 thrust axis determined from energy clusters in the calorimeter must
 have $\left|\cos\theta\right|<0.85$.
 within the acceptance of the
 vertex detector.
 These requirements yield a sample
 of $\sim 267,500$ hadronic $Z^0$ decays for the 1997-98 datset.

  Good quality tracks used for vertex finding must have a CDC hit
   at a radius$<$39~cm, and have $\geq$23 hits to insure that
the lever arm provided by the CDC is appreciable.
The CDC tracks must have $p_T>$250~MeV/$c\:$ and 
extrapolate to within 1~cm of the IP in $r\phi$
and within 1.5~cm in
$z$ to eliminate tracks which arise 
from interaction with the detector material.
The fit of the track must satisfy $\chi^2/$d.o.f.$<8$.
At least two good VXD3 links are required, and the combined
CDC/VXD fit must also satisfy $\chi^2/$d.o.f.$<8$.

     The topological vertex reconstruction is applied separately to
 the tracks in each hemisphere (defined with respect to the 
 event thrust axis).
 The vertexing algorithm is described in detail
 in Ref. \cite{zvnim} and summarized here.
    The vertices are reconstructed in 3-D coordinate space
  by defining a vertex function $V({\bf r})$ at each
  position ${\bf r}$. The helix
 parameters for each track $i$ are used to describe
  the 3-D track trajectory as a Gaussian tube $f_i({\bf r})$, where the
  width of the tube is the uncertainty in the measured track location
  close to the IP.
  A function $f_0({\bf r})$ is used to describe the location and
  uncertainty of the IP.
 $V({\bf r})$ is defined as a function of $f_0({\bf r})$ and
  the $f_i({\bf r})$ such that it is
   small in regions where fewer than two tracks (required
  for a vertex) have significant $f_i({\bf r})$, and large in
 regions of high track multiplicity.
  Maxima are found in $V({\bf r})$ and
  clustered into resolved spatial regions.
  Tracks are associated with these regions to form a set of
  topological vertices.

 The efficiency for reconstructing at least one secondary vertex
 in a $b$ hemisphere is $\sim 67$\% for VXD3.
 The efficiency
 falls at shorter decay length as it becomes harder to resolve
 the secondary vertex from the IP.
 For  hemispheres containing secondary vertices,
 the `seed' vertex is chosen to be the one
  with the highest $V({\bf r})$ value.
  Vertices consistent with a $K^0_s \to \pi^+\pi^-$ decay, within 
  a 14~MeV invariant mass window,
  are excluded from the seed
 vertex selection and the two tracks are discarded.

 A vertex axis is formed by a straight line joining the IP
  to the seed vertex.
 The 3-D distance of closest approach of a track to the vertex axis, T,
  and the distance from the IP along the vertex
 axis to this point, L, are calculated for all quality tracks.
  Monte Carlo studies show that tracks which are not directly associated
  with the seed vertex
  but which pass T$<0.1$~cm and L$/$D$>0.3$ (where D is the distance
  from the IP to the seed vertex) are more
 likely to have been produced by the $B$ decay sequence than
  to have an alternative origin.
 Hence such tracks are added to the set of
  tracks in the seed vertex to form the candidate $B$ decay vertex,
  containing tracks from both the $B$ and cascade
  $D$ decays. 
 This set of tracks is fitted 
 to a common vertex. If the probability of the vertex fit
  is greater than 5\% the $B$ decay location is taken to be the
   fitted vertex location.
 Since the vertex
  includes tracks form both the $B$ and cascade charm decay points
 the fit probability distribution is not flat and $\sim65$\%
  of vertices reconstructed have a fit probability
  below 5\%.
 The reconstruction of the $B$ decay location in this case is improved
 by dividing the tracks (if there are at least three) into
 two `sub-vertices' by  
  selecting the
  combination of tracks with maximum two sub-vertex fit probability.
  The $B$ decay location is given by the sub-vertex closest to the IP,
 reducing systematic dependence upon the physics of
   the $B \to D$ decay.   
    The distance
  from the IP to the $B$ decay location 
  is the reconstructed decay length.
 Since the purity of the $B$ charge
 reconstruction is lower for decays close to the IP, where tracks are
 more likely to be wrongly assigned, decay lengths are required
 to be $>1$~mm. To avoid using the vertices of tracks originating from
 interactions with the detector material, the distance of the decay vertex
 from the beamline is required to be $<22$~mm,
 i.e. more than $1$~mm inside the SLC beampipe.

\begin{figure}[htb]
\centering
\epsfysize12cm
\leavevmode
\epsffile[20 180 530 660]{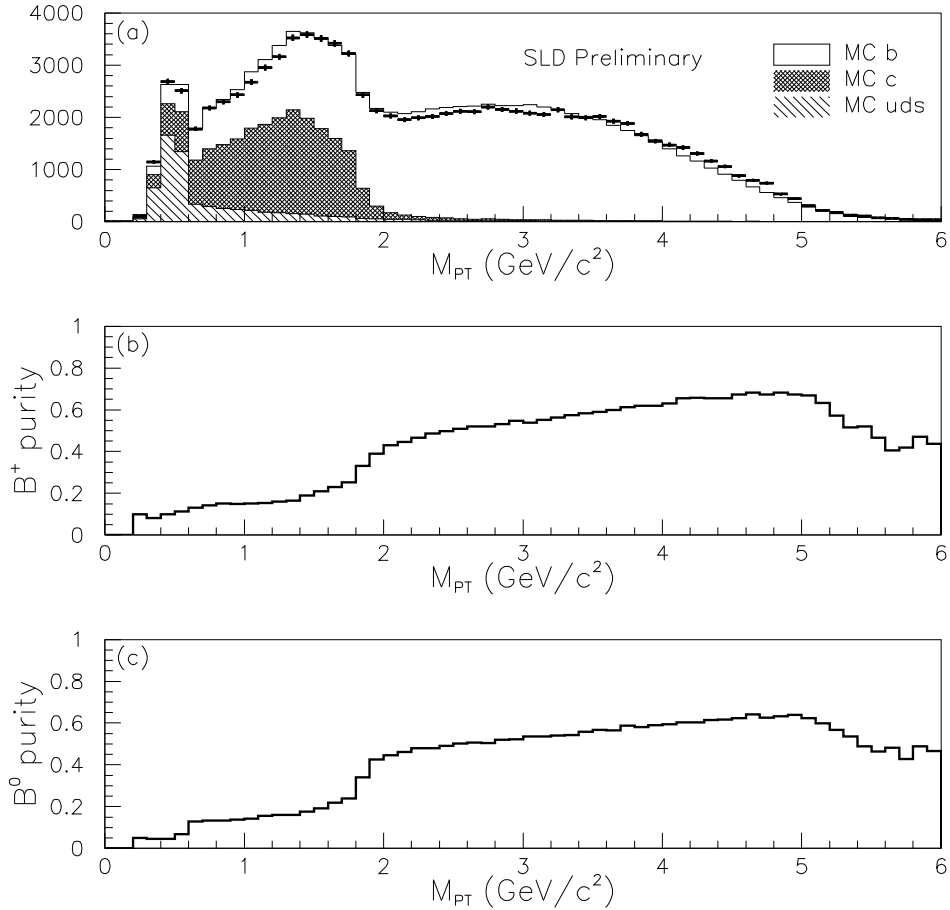}
\caption{ (a) $M_{P_T}$ of reconstructed vertex for 1997-98 
 data (points) and Monte Carlo (histogram), (b) $B^+$ fraction
 in the charged sample, (c) $B^0$ fraction in the neutral sample.}  
\label{mspt}
\end{figure}

 The mass $M$ of the reconstructed vertex is calculated by assuming each
 track has the mass of a pion.
 The transverse component $P_T$ of the total momentum of vertex tracks
 relative to the
 vertex axis is calculated in order to determine the $P_T$ corrected mass:
              
\begin{equation}
    M_{P_T}  =   \sqrt{M^2 + P^{2}_{T}} + | P_{T} |.
\label{mpt}
\end{equation}

 This quantity is
 the minimum mass the decaying hadron could have in order to
 produce a vertex with the quantities $M$ and $P_T$.
 The direction of the vertex axis is varied within the $1\sigma$ limits
 constraining the axis at the measured IP and reconstructed seed vertex
 such that the $P_{T}$ is minimized within this variation. This procedure
  prevents non-$B$ background vertices acquiring a high $M_{P_T}$ due to
  a fluctuation in the measured $P_{T}$. The accurate 3-D vertexing and
  precisely measured IP at SLD allow significant gain in the $b$-tag
  efficiency with high purity using this technique \cite{rbprl}.

    A comparison of the distribution of $M_{P_T}$
in data and Monte Carlo is
shown in Fig.~\ref{mspt}(a). 
 Vertices from $K^{0}_{s}$ decays surviving the $K^{0}_{s}$ rejection
 can be seen around $0.5$~GeV/$c^2$.
This figure shows that a large
 fraction of the charm and light flavor
contamination in the sample is eliminated by
requiring $M_{P_T} > 2$~GeV/$c^2$.
 It is also required that $M_{P_T} < 5.2$~GeV/$c^2$  since vertices
 with  $M_{P_T}$ greater than the $B$ meson mass are likely to contain
 background tracks.
These cuts yield a sample with $b$ hemisphere purity of $\sim$ 98\% with an
efficiency of 38\%. The average $B$ decay vertex multiplicity is 5.0 tracks.

  To improve the $B$ hadron charge reconstruction,
 tracks which
  fail the initial selection but have $p_T>$ 200 MeV/$c$ and
  $\sqrt{\sigma_{r\phi}^2 + \sigma_{rz}^2}<700\,\mu$m,
 where $\sigma_{r\phi}$ ($\sigma_{rz}$)
 is the uncertainty in the track position
 in the $r\phi$ ($rz$) plane
 close to the IP, are
  considered as decay track candidates.
    The charge of these tracks which
 pass the cuts
 T$<\,0.1$~cm and L$/$D$>0.3$ is added to the $B$ decay charge. On 
 average, 0.4 tracks pass these criteria in
 $b$ hemispheres. These lower quality tracks are
 used only to improve the charge
  reconstruction.

\begin{figure}[htb]
\centering
\epsfysize9cm
\leavevmode
\epsffile[0 165 540 670]{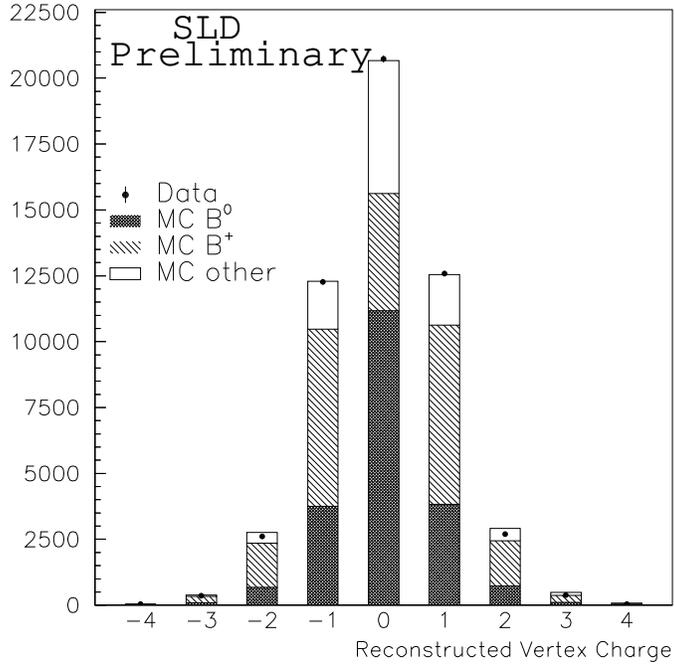}
\caption{Reconstructed vertex charge for 
 data (points) and Monte Carlo (histogram)
 for the 1997-98 dataset.}  
\label{char9798}
\end{figure}

 Fig.~\ref{char9798} shows a comparison of the reconstructed
charge between data and Monte Carlo for the 1997-1998
dataset. At this stage the charged sample 
consists of 30903 vertices with vertex
  charge equal to $\pm $ 1,2 or 3,
  while the neutral sample consists
  of 20731 vertices with charge equal to 0.
Monte Carlo studies
indicate that 
the charged sample is 97.0\% pure in $B$ hadrons
consisting of
57.2\% $B^+$, 32.0\% $B^0$,
 8.1\% $B_s^0$, and 4.5\% $B$ baryons.
(Charge conjugation is implied throughout this paper with the
 exception of the notation $B_+$ and $B_-$ introduced later to
 distinguish the charged mesons $B(\overline{b}u)$ and
 $B(b\overline{u})$ and respectively.) 
 Similarly, the neutral sample is 98.3\% pure in $B$ hadrons
consisting of 21.9\% $B^+$, 55.0\% $B^0$,
 15.6\% $B_s^0$ and 7.6\% $B$ baryons.
 The statistical precision of the measurement depends on the
 separation between the $B^+$ and $B^0$ in these samples.

 The lifetime measurement relies on the ability to separate
  $B^+$ and $B^0$ decays by making use of the  vertex charge.
   Monte Carlo studies show that
 the purity of the charge reconstruction
 is more likely to be eroded by losing
 tracks from the $B$ decay chain through track selection
 inefficiencies and track mis-assignment than by gaining mis-assigned
 tracks originating from the primary or other background to the $B$
 decay.
 Furthermore, the decays which are missing some $B$ tracks
  tend to have
 lower vertex mass as well as lower charge purity.
 Fig.~\ref{mspt}(b) and (c) show the fraction of $B^+$ decays in the 
 charged sample and the fraction of $B^0$ decays in the neutral sample
 respectively for MC. 
 The probability for the vertex to originate from a positively charged,
 neutral or negatively charged $B$ meson is denoted as $P_c(B_+)$,
 $P_c(B^0)$ and $P_c(B_-)$ respectively, normalized such that
 $P_c(B_+) + P_c(B^0) + P_c(B_-) = 1$.
  Vertices are given a weight, $w$, according
  to their analyzing
   power for separating $B^+$ and $B^0$ decays:
  $w = \left| 2P_c(B^+)  - 1 \right|$, where
   $P_c(B^+) = P_c(B_+) + P_c(B_-)$.
  The probabilities, and hence the weight,
  are a function of $M_{P_{T}}$.
  
\begin{figure}[htb]
\centering
\epsfysize10cm
\leavevmode
\epsffile[60 140 525 660]{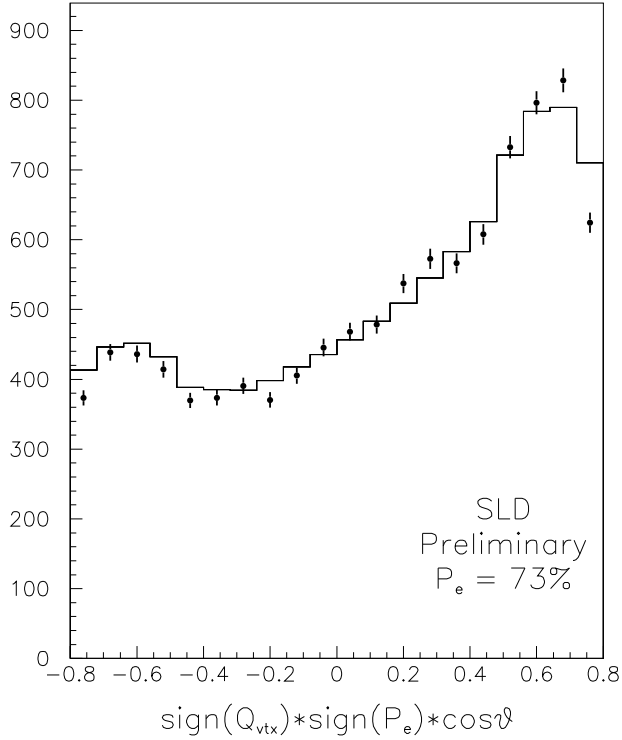}
\caption{Distribution of 
 cos$\theta$ signed by ($P_e \times$ vertex charge) for 
 1997-8 data (points) and Monte Carlo (histogram).}  
\label{pol}
\end{figure}

  The charge reconstruction in Fig.~\ref{char9798} shows good agreement
  between data and MC. A further check is made using the SLC electron
  beam polarization.
The polarized forward-backward asymmetry ${A}_{FB}(P_e,\cos\theta)$
  can be described by

\begin{equation}
 {A}_{FB}(P_e,\cos\theta) = 2 A_b~{{A_e - P_e}\over{1 - A_e P_e}}
        ~{{\cos\theta}\over{1+\cos^2\theta}}~,
\label{afb}
\end{equation}
where $A_b = 0.94$ and $A_e = 0.155$ (Standard Model values),
$P_e$ is the electron beam longitudinal polarization,
and $\theta$ is the angle between the
thrust axis and the electron beam direction (the thrust axis is signed
such that it points in the same hemisphere as the reconstructed vertex).
 Using negative (positive) vertex charge, with vertices weighted
 by the $M_{P_T}$ dependent analyzing power, to tag the
  $b$ ($\overline{b}$)  quark flavor the resulting forward backward
  asymmetry is sensitive
 to the accuracy of the vertex charge
  reconstruction.
   Good agreement between data and MC can be seen in Fig.~\ref{pol} 
  for the 1997-8 data, indicating that the MC
 adequately reproduces the charge reconstruction
 purity of the data. (Random vertex charge assignment
  would generate distributions with no signed $\cos\theta$ asymmetry.) 

The polarized forward-backward asymmetry
 can be used to tag the initial state $b$ or $\overline{b}$
 flavor of the hemisphere.
 The initial state $b/\overline{b}$ tag is also used to enhance
 the charged sample purity by giving a higher (lower) weight to
 the $B^+$ hypothesis if the vertex charge agrees 
 (disagrees) with the $b/\overline{b}$ tag. 
The probability for correctly tagging a $b$ quark at production
 using the $e^-$ beam polarization
is expressed as

\begin{equation}
P_A(b) = {{1 + A_{FB}(P_e,\cos\theta)}\over{2}}~.
\label{pa}
\end{equation}                      
A jet charge technique is used in addition to the polarized
forward-backward asymmetry. For this tag, tracks in the hemisphere
opposite that of the reconstructed vertex are selected. These tracks
are required to have momentum transverse to the beam axis
$p_\perp > 150$ MeV/c, total momentum $p < 50$ GeV/c, impact parameter
in the plane perpendicular to the beam axis $\delta < 2$ cm,
distance between the primary vertex and the track at the point of
closest approach along the beam axis $\Delta z < 10$ cm, and
$|\cos\theta| < 0.87$.
With these tracks, an opposite hemisphere momentum-weighted track
charge is defined as

\begin{equation}
Q_{opp} = \sum_i q_i \left|\vec{p}_i \cdot \hat{T}\right|^\kappa~,
\label{qopp}
\end{equation}
where $q_i$ is the electric charge of track $i$, $\vec{p}_i$ its momentum
vector, $\hat{T}$ is the thrust axis direction, and $\kappa$ is
a coefficient chosen to be 0.5 to maximize the separation between $b$ and
$\overline{b}$ quarks.
The probability for correctly tagging a $b$ quark in the initial state
of the vertex hemisphere can be parameterized as

\begin{equation}
P_Q(b) = {{1}\over{1 + e^{\alpha Q_{opp}}}}~,
\label{pq}
\end{equation}
where the coefficient $\alpha = -0.27$ as determined using the
Monte Carlo simulation. This technique is independent of the polarized
forward-backward asymmetry tag.
 The average purity of the $b/\overline{b}$ tag is $\sim72$\% using
 the forward backward asymmetry
  ($\left|P_e\right| = 73$\%) and 67\% for
 the jet charge technique.
   
The two initial state tags can be combined to form an overall initial
state tag with $b/\overline{b}$ quark probability
 $P_{i}(b)/P_{i}(\overline{b})$ (a function of $P_e$, $\cos\theta$
 and $Q_{opp}$).
 This probability is then combined with that obtained from the 
 vertex charge reconstruction (a function of $M_{P_T}$) to 
 determine the overall probability of a $B^+$  or $B^0$ decay:

\begin{equation}
P(B^+) = {{P_c(B_+)P_i(\overline{b}) + P_c(B_-)P_i(b)}\over
    {P_c(B_+)P_i(\overline{b}) + P_c(B_-)P_i(b) + 0.5\times P_c(B^0)}}
\label{pbp}
\end{equation}

  where $B_+$ and $B_-$ denote the positively and negatively charged $B$ mesons
  separately
  and $P(B^0) = 1 - P(B^+)$. 
 If the probability $P(B^+)>P(B^0)$ the vertex is classified 
 as charged, otherwise it is added to the neutral sample. In 
  either case it is weighted by the analyzing power 
   $w = \left| 2P(B^+) - 1 \right|$.
 Including the initial state tag information enhances the 
 statistical power of the analysis by $\sim$20\%.

\begin{figure}[htb]
\centering
\epsfysize6cm
\leavevmode
\epsffile[60 400 520 660]{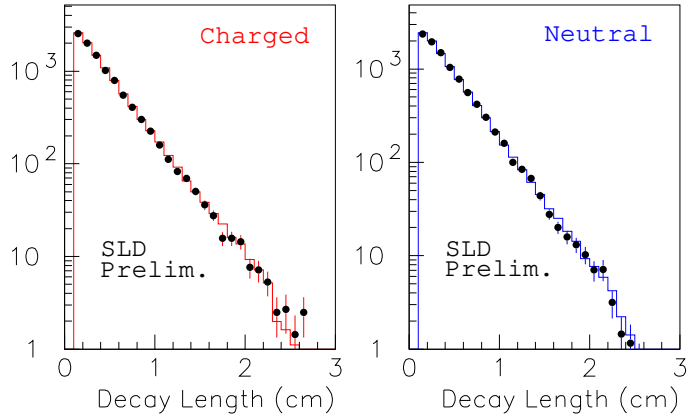}
\caption{Decay length distributions for the 1997-8
  data (points) and best fit Monte Carlo (histogram). }
\label{dklen}
\end{figure}

The $B^+$ and $B^0$ lifetimes are extracted from the decay length
distributions of the $B$ vertices in the charged and neutral
samples 
using a binned $\chi^2$ fit.
These distributions are fitted
simultaneously to determine the $B^+$ and $B^0$ lifetimes.
For each set of parameter values, Monte Carlo decay length distributions
are obtained by reweighting entries from generated $B^+$ and $B^0$
decays in the original Monte Carlo decay length
distributions with
$W(t,\tau)= \left( \frac{1}{\tau} ~ e^{-t/\tau} \right)/
 \left( \frac{1}{\tau_{gen}} ~ e^{-t/\tau_{gen}} \right)$,
where $\tau$ is the desired $B^+$ or $B^0$ lifetime,
$\tau_{gen}$ is the average lifetime generated in the Monte Carlo,
and $t$ is the proper time of each decay.
The fit then compares the decay length distributions
from the data with the reweighted Monte Carlo distributions.
 Fig.~\ref{dklen} shows the reconstructed decay length for data
 and best fit MC for the 1997-8 data and the charged and 
 neutral samples. (The fits are made using histograms in which
 the bin size increases with decay length such that the
 number of entries per bin remains approximately constant.)
 The  fit to the 1997-8 data sample yields
 lifetimes of
   $\tau_{B^+}=1.657\pm0.023$ ps and
   $\tau_{B^0}=1.578\pm0.024$ ps,
   with a ratio of $\tau_{B^+}/\tau_{B^0}$
   $ = 1.049^{+0.028}_{-0.027}$ and
   a $\chi^2/$d.o.f. = 79.8/76. These results are then corrected to
   correspond to b-fragmentation with a mean scaled energy $<X_E>=0.714$ 
  \cite{danning}, a 
  $B_s^0$ lifetime of 1.49 ps, and a b-baryon fraction of 10.2\%
  \cite{pdg98}. The corrected results for 1997-98 are:
   $\tau_{B^+}=1.613\pm0.023$ ps and
   $\tau_{B^0}=1.565\pm0.024$ ps,
   with a ratio of $\tau_{B^+}/\tau_{B^0}$
   $ = 1.030^{+0.028}_{-0.027}$.

To account for a discrepancy between data and Monte Carlo
in the fraction of tracks passing the selection criteria,
a $\sim$4\% tracking efficiency correction 
with dependence on track momenta and angles
 is applied to the simulation \cite{rbrb}.
 The corrected Monte Carlo is used in the lifetime fits, with the 
 effect of the entire correction 
taken as the systematic error. 

The physics modeling systematic uncertainties were determined as follows.
The mean fragmentation energy $<\!\! x_E\!\!>$ of the $B$ hadron 
 \cite{danning} 
 and the shape of the $x_E$ distribution \cite{Bowler} were varied.
 Since the fragmentation is assumed to be identical for the $B^+$ and
 $B^0$ mesons, this uncertainty has little effect on the lifetime ratio.
The four branching fractions
for $B^+/B^0\rightarrow \overline{D^0}/D^- X$
were varied by twice the uncertainty given in Ref.~\cite{muheim} for
 $B\rightarrow \overline{D^0}/D^- X$.
The fraction of $B^+/B^0$ decays producing a $D\overline{D}$ pair was also
 varied.
The average $B^+$ and $B^0$ decay multiplicity was varied by
 $\pm 0.3$ tracks \cite{bmult} in an anticorrelated manner.
Uncertainties in the $B_s^0$ and $B$ baryon lifetimes and production
 fractions mostly affect the $B^0$ lifetime since 
the  neutral $B_s^0$ and $B$ baryon are a
 more significant background
  for the $B^0$ decays.
The systematic errors due to uncertainties in charmed meson decay
topology were estimated by changing the Monte Carlo
 $D$ decay charged multiplicity and $K^0$ production
according to the uncertainties in experimental measurements \cite{MK3DDECAY}.
The effect of varying the lifetime of charm hadrons
  ($D^+$, $D^0$, $D_s$, $\Lambda_c$), as well as their momentum spectra in
 the $B$ decay rest frame was found to be negligible.

 The fitting uncertainties were determined
by varying the bin size used in the decay length distributions,
and by modifying the cuts on the minimum decay length (no cut--2~mm)
  and maximum radius cuts
 (20~mm--no cut) 
used in the fit. Fit results are consistent within statistics
for these variations, but a systematic error is conservatively assigned
using the RMS variation of the results.

Table~\ref{systerrs} summarizes the systematic errors on the $B^+$ and
$B^0$ lifetimes and their ratio for the 1997-1998 dataset. 

\begin{table}[th]
\begin{center}
\caption{\label{systerrs} Summary of systematic uncertainties in
         the $B^+$ and $B^0$ lifetimes and their ratio
          for the 1997-98 dataset.}

\begin{tabular}{lcccc}
\\
\hline
Systematic Error  & & $\Delta\tau_{B^+}$
                    & $\Delta\tau_{B^0}$
                    & $\Delta\frac{\tau_{B^+}}{\tau_{B^0}}$  \\
  & & (ps) & (ps) & \\
   \hline
 & \multicolumn{3}{c}{Detector Modeling} \\ \hline
 Tracking efficiency& & 0.005  & 0.011 & 0.008 \\
 Tracking resolution& & 0.003  & 0.003 & 0.003 \\
\hline
 & \multicolumn{3}{c}{Physics Modeling} \\ \hline
$b$ fragmentation  & 0.714 $\pm$ 0.008 & 0.025 & 0.030 & $<$0.003 \\
     & $x_E$ shape  & 0.011 & 0.009 & $<$0.003  \\
BR($B \to DX$)   &  & 0.005    & 0.009  & 0.007    \\
BR($B \to D\overline{D}X$)   & 0.19 $\pm$ 0.05 & 0.015 & 0.014 & 0.016 \\
$B$ decay multiplicity & 5.3 $\pm$ 0.3 & $<$0.003 & $<$0.003 & $<$0.003    \\
$B_s^0$ fraction  & 0.115 $\pm$ 0.020 & 0.003    &  $<$0.003  & $<$0.003  \\
$B$ baryon fraction & 0.102 $\pm$ 0.020 & 0.004    & 0.017  & 0.008  \\
$B_s^0$ lifetime & 1.49 $\pm$ 0.06 ps  & $<$0.003    & 0.014  & 0.009  \\
$B$ baryon lifetime & 1.22 $\pm$ 0.05 ps & $<$0.003  & 0.005  & 0.003  \\
$D$ decay multiplicity&  & 0.003 & 0.005 & 0.006 \\
$D$ decay $K^0$ yield& & $<$0.003   & 0.008   & 0.004   \\
\hline
 & \multicolumn{3}{c}{Monte Carlo and Fitting} \\ \hline
Fitting systematics & & 0.010    & 0.009 & 0.006  \\
MC statistics      & & 0.006    & 0.006 & 0.007   \\
\hline
TOTAL            &   & 0.033    & 0.044 & 0.026     \\
\hline
\end{tabular}
\end{center}
\end{table}

Finally, the 1993-95 and 1996 results \cite{9396results} are 
then combined with these
more recent results, taking into account correlated uncertainties.
For the earlier data sets, the results have been adjusted so that
their assumed values for the b-baryon fraction and the mean of the 
fragmentation function agrees with this current analysis. 
The  combination of the 1993-5, 1996, and 1997-98 data samples yield
lifetimes of
   $\tau_{B^+}=1.623\pm0.020$ ps and
   $\tau_{B^0}=1.585\pm0.031$ ps,
   with a ratio of $\tau_{B^+}/\tau_{B^0}$
   $ = 1.037^{+0.025}_{-0.024}$. 
The combined systematics for the 1993-98 data are shown in Table~\ref{sys9398}.

\begin{table}[th]
\begin{center}
\caption{\label{sys9398} Summary of systematic uncertainties in
         the $B^+$ and $B^0$ lifetimes and their ratio 
         for the combined 1993-8 data.}

\begin{tabular}{lcccc}
\\
\hline
Systematic Error  & & $\Delta\tau_{B^+}$
                    & $\Delta\tau_{B^0}$
                    & $\Delta\frac{\tau_{B^+}}{\tau_{B^0}}$  \\
  & & (ps) & (ps) & \\
   \hline
 & \multicolumn{3}{c}{Detector Modeling} \\ \hline
 Tracking efficiency& & 0.004  & 0.004 & 0.005 \\
 Tracking resolution& & 0.003  & 0.003 & 0.003 \\
\hline
 & \multicolumn{3}{c}{Physics Modeling} \\ \hline
$b$ fragmentation  & 0.714 $\pm$ 0.008 & 0.025 & 0.028 & .004 \\
     & $x_E$ shape  & 0.011 & 0.010 & $<$0.003  \\
BR($B \to DX$)   &  & 0.005    & 0.009  & 0.007    \\
BR($B \to D\overline{D}X$)   & 0.19 $\pm$ 0.05 & 0.013 & 0.012 & 0.014 \\
$B$ decay multiplicity & 5.3 $\pm$ 0.3 &$<$0.003    &$<$0.003  &$<$0.003    \\
$B_s^0$ fraction  & 0.115 $\pm$ 0.020 & 0.005    &  $<$0.003  & $<$0.003  \\
$B$ baryon fraction & 0.102 $\pm$ 0.020 & 0.005    & 0.017  & 0.008  \\
$B_s^0$ lifetime & 1.49 $\pm$ 0.06 ps  & $<$0.003    & 0.014  & 0.009  \\
$B$ baryon lifetime & 1.22 $\pm$ 0.05 ps & $<$0.003  & 0.005  & 0.003  \\
$D$ decay multiplicity&  & 0.005 & 0.006 & 0.008 \\
$D$ decay $K^0$ yield& & 0.003   & 0.009   & 0.006   \\
\hline
 & \multicolumn{3}{c}{Monte Carlo and Fitting} \\ \hline
Fitting systematics & & 0.008    & 0.007 & 0.005   \\
MC statistics      & & 0.005    & 0.005 & 0.006   \\
\hline
TOTAL            &   & 0.034    & 0.043 & 0.024     \\
\hline
\end{tabular}
\end{center}
\end{table}

 In summary, from the entire 550,000 $Z^0$ decays
 collected by SLD between 1993 and 1998,
 the $B^+$ and $B^0$ lifetimes have been measured using an
 inclusive
 topological technique. The analysis of the 1997-98 dataset of 350,000
 decays, discussed above, isolates 
 51634 $B$ hadron candidates with good
 charge purity enhanced by the vertex mass, $e^-$ beam polarization
 and opposite hemisphere jet charge information. These results have
 been combined with the earlier 1993-96 measurements to yield the 
 following preliminary results:

\begin{eqnarray}
\tau_{B^+} & = &
  1.623\pm0.020(\mbox{stat})\pm0.034(\mbox{syst})\mbox{~ps}, \nonumber \\
\tau_{B^0} & = &
  1.585\pm0.021(\mbox{stat})\pm0.043(\mbox{syst})\mbox{~ps}, \nonumber \\
\frac{\tau_{B^+}}{\tau_{B^0}} & = & 
  1.037\pm^{0.025}_{0.024}(\mbox{stat})\pm0.024(\mbox{syst}). \nonumber
\end{eqnarray}

 These results are consistent with the
expectation that the $B^+$ lifetime is up to
 10\% greater than that of the $B^0$  and have the best
  statistical accuracy among current measurements \cite{world}.

\noindent

        We thank the personnel of the SLAC accelerator department and
the technical staffs of our collaborating institutions for their outstanding
efforts.

\pagebreak

%
%
%
\section*{$^{**}$ List of Authors}

\begin{center}
\def\iADEL{$^{(1)}$}
\def\iAOMORI{$^{(2)}$}
\def\iBOLO{$^{(3)}$}
\def\iBRI{$^{(4)}$}
\def\iBRUN{$^{(5)}$}
\def\iBU{$^{(6)}$}
\def\iCINC{$^{(7)}$}
\def\iCOLO{$^{(8)}$}
\def\iCOLU{$^{(9)}$}
\def\iCSU{$^{(10)}$}
\def\iFERR{$^{(11)}$}
\def\iFRAS{$^{(12)}$}
\def\iILLI{$^{(13)}$}
\def\iJHU{$^{(14)}$}
\def\iLBL{$^{(15)}$}
\def\iLTU{$^{(16)}$}
\def\iMASS{$^{(17)}$}
\def\iMISSI{$^{(18)}$}
\def\iMIT{$^{(19)}$}
\def\iMOSCOW{$^{(20)}$}
\def\iNAGO{$^{(21)}$}
\def\iOREG{$^{(22)}$}
\def\iOXF{$^{(23)}$}
\def\iPADO{$^{(24)}$}
\def\iPERU{$^{(25)}$}
\def\iPISA{$^{(26)}$}
\def\iRAL{$^{(27)}$}
\def\iRUTG{$^{(28)}$}
\def\iSLAC{$^{(29)}$}
\def\iSOGA{$^{(30)}$}
\def\iSOONG{$^{(31)}$}
\def\iTENN{$^{(32)}$}
\def\iTOHO{$^{(33)}$}
\def\iUCSB{$^{(34)}$}
\def\iUCSC{$^{(35)}$}
\def\iUVIC{$^{(36)}$}
\def\iVAND{$^{(37)}$}
\def\iWASH{$^{(38)}$}
\def\iWISC{$^{(39)}$}
\def\iYALE{$^{(40)}$}

  \baselineskip=.75\baselineskip  
\mbox{Kenji  Abe\unskip,\iNAGO}
\mbox{Koya Abe\unskip,\iTOHO}
\mbox{T. Abe\unskip,\iSLAC}
\mbox{I. Adam\unskip,\iSLAC}
\mbox{T.  Akagi\unskip,\iSLAC}
\mbox{H. Akimoto\unskip,\iSLAC}
\mbox{N.J. Allen\unskip,\iBRUN}
\mbox{W.W. Ash\unskip,\iSLAC}
\mbox{D. Aston\unskip,\iSLAC}
\mbox{K.G. Baird\unskip,\iMASS}
\mbox{C. Baltay\unskip,\iYALE}
\mbox{H.R. Band\unskip,\iWISC}
\mbox{M.B. Barakat\unskip,\iLTU}
\mbox{O. Bardon\unskip,\iMIT}
\mbox{T.L. Barklow\unskip,\iSLAC}
\mbox{G.L. Bashindzhagyan\unskip,\iMOSCOW}
\mbox{J.M. Bauer\unskip,\iMISSI}
\mbox{G. Bellodi\unskip,\iOXF}
\mbox{A.C. Benvenuti\unskip,\iBOLO}
\mbox{G.M. Bilei\unskip,\iPERU}
\mbox{D. Bisello\unskip,\iPADO}
\mbox{G. Blaylock\unskip,\iMASS}
\mbox{J.R. Bogart\unskip,\iSLAC}
\mbox{G.R. Bower\unskip,\iSLAC}
\mbox{J.E. Brau\unskip,\iOREG}
\mbox{M. Breidenbach\unskip,\iSLAC}
\mbox{W.M. Bugg\unskip,\iTENN}
\mbox{D. Burke\unskip,\iSLAC}
\mbox{T.H. Burnett\unskip,\iWASH}
\mbox{P.N. Burrows\unskip,\iOXF}
\mbox{R.M. Byrne\unskip,\iMIT}
\mbox{A. Calcaterra\unskip,\iFRAS}
\mbox{D. Calloway\unskip,\iSLAC}
\mbox{B. Camanzi\unskip,\iFERR}
\mbox{M. Carpinelli\unskip,\iPISA}
\mbox{R. Cassell\unskip,\iSLAC}
\mbox{R. Castaldi\unskip,\iPISA}
\mbox{A. Castro\unskip,\iPADO}
\mbox{M. Cavalli-Sforza\unskip,\iUCSC}
\mbox{A. Chou\unskip,\iSLAC}
\mbox{E. Church\unskip,\iWASH}
\mbox{H.O. Cohn\unskip,\iTENN}
\mbox{J.A. Coller\unskip,\iBU}
\mbox{M.R. Convery\unskip,\iSLAC}
\mbox{V. Cook\unskip,\iWASH}
\mbox{R.F. Cowan\unskip,\iMIT}
\mbox{D.G. Coyne\unskip,\iUCSC}
\mbox{G. Crawford\unskip,\iSLAC}
\mbox{C.J.S. Damerell\unskip,\iRAL}
\mbox{M.N. Danielson\unskip,\iCOLO}
\mbox{M. Daoudi\unskip,\iSLAC}
\mbox{N. de Groot\unskip,\iBRI}
\mbox{R. Dell'Orso\unskip,\iPERU}
\mbox{P.J. Dervan\unskip,\iBRUN}
\mbox{R. de Sangro\unskip,\iFRAS}
\mbox{M. Dima\unskip,\iCSU}
\mbox{D.N. Dong\unskip,\iMIT}
\mbox{M. Doser\unskip,\iSLAC}
\mbox{R. Dubois\unskip,\iSLAC}
\mbox{B.I. Eisenstein\unskip,\iILLI}
\mbox{I.Erofeeva\unskip,\iMOSCOW}
\mbox{V. Eschenburg\unskip,\iMISSI}
\mbox{E. Etzion\unskip,\iWISC}
\mbox{S. Fahey\unskip,\iCOLO}
\mbox{D. Falciai\unskip,\iFRAS}
\mbox{C. Fan\unskip,\iCOLO}
\mbox{J.P. Fernandez\unskip,\iUCSC}
\mbox{M.J. Fero\unskip,\iMIT}
\mbox{K. Flood\unskip,\iMASS}
\mbox{R. Frey\unskip,\iOREG}
\mbox{J. Gifford\unskip,\iUVIC}
\mbox{T. Gillman\unskip,\iRAL}
\mbox{G. Gladding\unskip,\iILLI}
\mbox{S. Gonzalez\unskip,\iMIT}
\mbox{E.R. Goodman\unskip,\iCOLO}
\mbox{E.L. Hart\unskip,\iTENN}
\mbox{J.L. Harton\unskip,\iCSU}
\mbox{K. Hasuko\unskip,\iTOHO}
\mbox{S.J. Hedges\unskip,\iBU}
\mbox{S.S. Hertzbach\unskip,\iMASS}
\mbox{M.D. Hildreth\unskip,\iSLAC}
\mbox{J. Huber\unskip,\iOREG}
\mbox{M.E. Huffer\unskip,\iSLAC}
\mbox{E.W. Hughes\unskip,\iSLAC}
\mbox{X. Huynh\unskip,\iSLAC}
\mbox{H. Hwang\unskip,\iOREG}
\mbox{M. Iwasaki\unskip,\iOREG}
\mbox{D.J. Jackson\unskip,\iRAL}
\mbox{P. Jacques\unskip,\iRUTG}
\mbox{J.A. Jaros\unskip,\iSLAC}
\mbox{Z.Y. Jiang\unskip,\iSLAC}
\mbox{A.S. Johnson\unskip,\iSLAC}
\mbox{J.R. Johnson\unskip,\iWISC}
\mbox{R.A. Johnson\unskip,\iCINC}
\mbox{T. Junk\unskip,\iSLAC}
\mbox{R. Kajikawa\unskip,\iNAGO}
\mbox{M. Kalelkar\unskip,\iRUTG}
\mbox{Y. Kamyshkov\unskip,\iTENN}
\mbox{H.J. Kang\unskip,\iRUTG}
\mbox{I. Karliner\unskip,\iILLI}
\mbox{H. Kawahara\unskip,\iSLAC}
\mbox{Y.D. Kim\unskip,\iSOGA}
\mbox{M.E. King\unskip,\iSLAC}
\mbox{R. King\unskip,\iSLAC}
\mbox{R.R. Kofler\unskip,\iMASS}
\mbox{N.M. Krishna\unskip,\iCOLO}
\mbox{R.S. Kroeger\unskip,\iMISSI}
\mbox{M. Langston\unskip,\iOREG}
\mbox{A. Lath\unskip,\iMIT}
\mbox{D.W.G. Leith\unskip,\iSLAC}
\mbox{V. Lia\unskip,\iMIT}
\mbox{C.Lin\unskip,\iMASS}
\mbox{M.X. Liu\unskip,\iYALE}
\mbox{X. Liu\unskip,\iUCSC}
\mbox{M. Loreti\unskip,\iPADO}
\mbox{A. Lu\unskip,\iUCSB}
\mbox{H.L. Lynch\unskip,\iSLAC}
\mbox{J. Ma\unskip,\iWASH}
\mbox{M. Mahjouri\unskip,\iMIT}
\mbox{G. Mancinelli\unskip,\iRUTG}
\mbox{S. Manly\unskip,\iYALE}
\mbox{G. Mantovani\unskip,\iPERU}
\mbox{T.W. Markiewicz\unskip,\iSLAC}
\mbox{T. Maruyama\unskip,\iSLAC}
\mbox{H. Masuda\unskip,\iSLAC}
\mbox{E. Mazzucato\unskip,\iFERR}
\mbox{A.K. McKemey\unskip,\iBRUN}
\mbox{B.T. Meadows\unskip,\iCINC}
\mbox{G. Menegatti\unskip,\iFERR}
\mbox{R. Messner\unskip,\iSLAC}
\mbox{P.M. Mockett\unskip,\iWASH}
\mbox{K.C. Moffeit\unskip,\iSLAC}
\mbox{T.B. Moore\unskip,\iYALE}
\mbox{M.Morii\unskip,\iSLAC}
\mbox{D. Muller\unskip,\iSLAC}
\mbox{V. Murzin\unskip,\iMOSCOW}
\mbox{T. Nagamine\unskip,\iTOHO}
\mbox{S. Narita\unskip,\iTOHO}
\mbox{U. Nauenberg\unskip,\iCOLO}
\mbox{H. Neal\unskip,\iSLAC}
\mbox{M. Nussbaum\unskip,\iCINC}
\mbox{N. Oishi\unskip,\iNAGO}
\mbox{D. Onoprienko\unskip,\iTENN}
\mbox{L.S. Osborne\unskip,\iMIT}
\mbox{R.S. Panvini\unskip,\iVAND}
\mbox{C.H. Park\unskip,\iSOONG}
\mbox{T.J. Pavel\unskip,\iSLAC}
\mbox{I. Peruzzi\unskip,\iFRAS}
\mbox{M. Piccolo\unskip,\iFRAS}
\mbox{L. Piemontese\unskip,\iFERR}
\mbox{K.T. Pitts\unskip,\iOREG}
\mbox{R.J. Plano\unskip,\iRUTG}
\mbox{R. Prepost\unskip,\iWISC}
\mbox{C.Y. Prescott\unskip,\iSLAC}
\mbox{G.D. Punkar\unskip,\iSLAC}
\mbox{J. Quigley\unskip,\iMIT}
\mbox{B.N. Ratcliff\unskip,\iSLAC}
\mbox{T.W. Reeves\unskip,\iVAND}
\mbox{J. Reidy\unskip,\iMISSI}
\mbox{P.L. Reinertsen\unskip,\iUCSC}
\mbox{P.E. Rensing\unskip,\iSLAC}
\mbox{L.S. Rochester\unskip,\iSLAC}
\mbox{P.C. Rowson\unskip,\iCOLU}
\mbox{J.J. Russell\unskip,\iSLAC}
\mbox{O.H. Saxton\unskip,\iSLAC}
\mbox{T. Schalk\unskip,\iUCSC}
\mbox{R.H. Schindler\unskip,\iSLAC}
\mbox{B.A. Schumm\unskip,\iUCSC}
\mbox{J. Schwiening\unskip,\iSLAC}
\mbox{S. Sen\unskip,\iYALE}
\mbox{V.V. Serbo\unskip,\iSLAC}
\mbox{M.H. Shaevitz\unskip,\iCOLU}
\mbox{J.T. Shank\unskip,\iBU}
\mbox{G. Shapiro\unskip,\iLBL}
\mbox{D.J. Sherden\unskip,\iSLAC}
\mbox{K.D. Shmakov\unskip,\iTENN}
\mbox{C. Simopoulos\unskip,\iSLAC}
\mbox{N.B. Sinev\unskip,\iOREG}
\mbox{S.R. Smith\unskip,\iSLAC}
\mbox{M.B. Smy\unskip,\iCSU}
\mbox{J.A. Snyder\unskip,\iYALE}
\mbox{H. Staengle\unskip,\iCSU}
\mbox{A. Stahl\unskip,\iSLAC}
\mbox{P. Stamer\unskip,\iRUTG}
\mbox{H. Steiner\unskip,\iLBL}
\mbox{R. Steiner\unskip,\iADEL}
\mbox{M.G. Strauss\unskip,\iMASS}
\mbox{D. Su\unskip,\iSLAC}
\mbox{F. Suekane\unskip,\iTOHO}
\mbox{A. Sugiyama\unskip,\iNAGO}
\mbox{S. Suzuki\unskip,\iNAGO}
\mbox{M. Swartz\unskip,\iJHU}
\mbox{A. Szumilo\unskip,\iWASH}
\mbox{T. Takahashi\unskip,\iSLAC}
\mbox{F.E. Taylor\unskip,\iMIT}
\mbox{J. Thom\unskip,\iSLAC}
\mbox{E. Torrence\unskip,\iMIT}
\mbox{N.K. Toumbas\unskip,\iSLAC}
\mbox{T. Usher\unskip,\iSLAC}
\mbox{C. Vannini\unskip,\iPISA}
\mbox{J. Va'vra\unskip,\iSLAC}
\mbox{E. Vella\unskip,\iSLAC}
\mbox{J.P. Venuti\unskip,\iVAND}
\mbox{R. Verdier\unskip,\iMIT}
\mbox{P.G. Verdini\unskip,\iPISA}
\mbox{D.L. Wagner\unskip,\iCOLO}
\mbox{S.R. Wagner\unskip,\iSLAC}
\mbox{A.P. Waite\unskip,\iSLAC}
\mbox{S. Walston\unskip,\iOREG}
\mbox{S.J. Watts\unskip,\iBRUN}
\mbox{A.W. Weidemann\unskip,\iTENN}
\mbox{E. R. Weiss\unskip,\iWASH}
\mbox{J.S. Whitaker\unskip,\iBU}
\mbox{S.L. White\unskip,\iTENN}
\mbox{F.J. Wickens\unskip,\iRAL}
\mbox{B. Williams\unskip,\iCOLO}
\mbox{D.C. Williams\unskip,\iMIT}
\mbox{S.H. Williams\unskip,\iSLAC}
\mbox{S. Willocq\unskip,\iMASS}
\mbox{R.J. Wilson\unskip,\iCSU}
\mbox{W.J. Wisniewski\unskip,\iSLAC}
\mbox{J. L. Wittlin\unskip,\iMASS}
\mbox{M. Woods\unskip,\iSLAC}
\mbox{G.B. Word\unskip,\iVAND}
\mbox{T.R. Wright\unskip,\iWISC}
\mbox{J. Wyss\unskip,\iPADO}
\mbox{R.K. Yamamoto\unskip,\iMIT}
\mbox{J.M. Yamartino\unskip,\iMIT}
\mbox{X. Yang\unskip,\iOREG}
\mbox{J. Yashima\unskip,\iTOHO}
\mbox{S.J. Yellin\unskip,\iUCSB}
\mbox{C.C. Young\unskip,\iSLAC}
\mbox{H. Yuta\unskip,\iAOMORI}
\mbox{G. Zapalac\unskip,\iWISC}
\mbox{R.W. Zdarko\unskip,\iSLAC}
\mbox{J. Zhou\unskip.\iOREG}

\it
  \vskip \baselineskip                   
  \centerline{(The SLD Collaboration)}   
  \vskip \baselineskip        
  \baselineskip=.75\baselineskip   
\iADEL
  Adelphi University, Garden City, New York 11530, \break
\iAOMORI
  Aomori University, Aomori , 030 Japan, \break
\iBOLO
  INFN Sezione di Bologna, I-40126, Bologna, Italy, \break
\iBRI
  University of Bristol, Bristol, U.K., \break
\iBRUN
  Brunel University, Uxbridge, Middlesex, UB8 3PH United Kingdom, \break
\iBU
  Boston University, Boston, Massachusetts 02215, \break
\iCINC
  University of Cincinnati, Cincinnati, Ohio 45221, \break
\iCOLO
  University of Colorado, Boulder, Colorado 80309, \break
\iCOLU
  Columbia University, New York, New York 10533, \break
\iCSU
  Colorado State University, Ft. Collins, Colorado 80523, \break
\iFERR
  INFN Sezione di Ferrara and Universita di Ferrara, I-44100 Ferrara, Italy, \break
\iFRAS
  INFN Lab. Nazionali di Frascati, I-00044 Frascati, Italy, \break
\iILLI
  University of Illinois, Urbana, Illinois 61801, \break
\iJHU
  Johns Hopkins University,  Baltimore, Maryland 21218-2686, \break
\iLBL
  Lawrence Berkeley Laboratory, University of California, Berkeley, California 94720, \break
\iLTU
  Louisiana Technical University, Ruston,Louisiana 71272, \break
\iMASS
  University of Massachusetts, Amherst, Massachusetts 01003, \break
\iMISSI
  University of Mississippi, University, Mississippi 38677, \break
\iMIT
  Massachusetts Institute of Technology, Cambridge, Massachusetts 02139, \break
\iMOSCOW
  Institute of Nuclear Physics, Moscow State University, 119899, Moscow Russia, \break
\iNAGO
  Nagoya University, Chikusa-ku, Nagoya, 464 Japan, \break
\iOREG
  University of Oregon, Eugene, Oregon 97403, \break
\iOXF
  Oxford University, Oxford, OX1 3RH, United Kingdom, \break
\iPADO
  INFN Sezione di Padova and Universita di Padova I-35100, Padova, Italy, \break
\iPERU
  INFN Sezione di Perugia and Universita di Perugia, I-06100 Perugia, Italy, \break
\iPISA
  INFN Sezione di Pisa and Universita di Pisa, I-56010 Pisa, Italy, \break
\iRAL
  Rutherford Appleton Laboratory, Chilton, Didcot, Oxon OX11 0QX United Kingdom, \break
\iRUTG
  Rutgers University, Piscataway, New Jersey 08855, \break
\iSLAC
  Stanford Linear Accelerator Center, Stanford University, Stanford, California 94309, \break
\iSOGA
  Sogang University, Seoul, Korea, \break
\iSOONG
  Soongsil University, Seoul, Korea 156-743, \break
\iTENN
  University of Tennessee, Knoxville, Tennessee 37996, \break
\iTOHO
  Tohoku University, Sendai 980, Japan, \break
\iUCSB
  University of California at Santa Barbara, Santa Barbara, California 93106, \break
\iUCSC
  University of California at Santa Cruz, Santa Cruz, California 95064, \break
\iUVIC
  University of Victoria, Victoria, British Columbia, Canada V8W 3P6, \break
\iVAND
  Vanderbilt University, Nashville,Tennessee 37235, \break
\iWASH
  University of Washington, Seattle, Washington 98105, \break
\iWISC
  University of Wisconsin, Madison,Wisconsin 53706, \break
\iYALE
  Yale University, New Haven, Connecticut 06511. \break

\rm
%

\end{center}

\enddocument